\documentclass[twocolumn,preprintnumbers,amssymb,prb,superscriptaddress,floatfix]{revtex4}
\usepackage{graphicx}
\usepackage{dcolumn}
\usepackage{bm}
\usepackage{amsmath}
\usepackage{latexsym}
\usepackage[dvips]{color}
\usepackage{psfrag}
\renewcommand{\wp}{\omega_{\rm p}}
\newcommand{\au}{{\mbox{$\;[$\,a.u.$]$}}}
\newcommand{\be}{\begin{equation}}
\newcommand{\ee}{\end{equation}}

\newcommand{\rd}{{\rm d}}
\newcommand{\vjel}{v_0}
\newcommand{\wl}{\omega_{\rm L}}
\begin{document}
\title{Optical response of two-dimensional electron fluids  beyond
the Kohn regime: strong non-parabolic confinement and intense laser
light}
\author{M. Santer}
\affiliation{Theoretical Quantum Dynamics, 
University of Freiburg, 79104 Freiburg, Germany}
\author{B. Mehlig}
\affiliation{School of Physics and Engineering Physics,
Gothenburg University/Chalmers, 412 96 Gothenburg, Sweden}
\author{M. Moseler$^1$}
\begin{abstract}
We investigate the linear and non-linear optical response 
of two-dimensional (2D) interacting electron fluids  confined by a
strong non-parabolic potential. We show that
such fluids  may exhibit
higher-harmonic spectra under realistic
experimental conditions. Higher harmonics arise as the electrons 
explore anharmonicities of the confinement potential
(electron-electron interactions reduce this non-linear effect).
This opens the possibility of controlling the optical
functionality of such systems by engineering the confinement
potential. Our results were obtained within 
time-dependent density-functional theory, employing
the adiabatic local-density approximation.
A classical hydrodynamical model is in good agreement with
the quantum-mechanical results.
\end{abstract}
\pacs{03.65.Sq, 73.20.Dx, 73.23.-b, 0545.Mt}
\maketitle
Confined, two-dimensional (2D) electronic quantum systems have been subject
to intense theoretical and experimental investigations
during the last two decades~\cite{and82}. Emerging from quantum-well
structures
or charge layers in modulation doped semiconductor interfaces,
they are nowadays routinely tailored into quantum dots~\cite{kou01,yoram}
or strips~\cite{ege90}.  Possible applications range
from single-electron transistors to  coherent, tunable light sources
for far-infrared  (FIR) spectroscopy~\cite{vos96}, a method which has
proven a powerful tool
for probing slow vibrational modes in molecular and condensed matter
systems.
In the light of this application and in order to  design future  2D THz
devices, a detailed understanding of linear -- and non-linear -- excitation
mechanisms in confined 2D electronic quantum systems is necessary.  

In  finite
2D systems (such as, e.g., quantum dots or quantum strips),
the linear optical response  depends
on the shape of the confinement potential $\vjel$.
Recent experimental results concern
parabolic or near-parabolic confinement potentials~\cite{kou01}
for which the so-called
harmonic-potential theorem (HPT) states that an external
dipole excitation can only couple to 
a rigid-shift mode (Kohn mode~\cite{koh61}) at frequency
$\sqrt{K/m^*}$
independently of the excitation strength~\cite{dob94}
($m^*$ is the effective 
electron mass and $K$ the curvature of $\vjel$).
In its original formulation, the HPT is a quantum-mechanical
theorem\cite{yip91,dob94}; it was shown\cite{scha96} to hold
in classical mechanics also.

In realistic, finite 2D quantum structures, the confinement potential
$\vjel$ is often
strongly modulated (e.g. by inhomogeneous charge 
distributions) exhibiting a pronounced anharmonicity~\cite{lie93}.
Nevertheless, in many experiments, the Kohn mode dominates
the response\cite{dem90}. This is because 
for weak external fields, and for low electron densities,
anharmonic regions of the confinement potential are 
hardly explored. Experimentally it is possible to overcome the HPT limitations
in at least two ways: either by increasing the density $\overline{n}$ of
conduction electrons or by increasing the intensity of
the laser light. This makes it possible to experimentally
investigate the hydrodynamics of the interacting electron fluid,
which is expected to reveal much more information
about the electron dynamics than the rigid-shift response in the Kohn regime.

However, in order to adequately describe the case of strong fields,
the non-linear response of the conduction
electrons must be considered. While 
the non-linear
response of atoms~\cite{hui92,chu01}, molecules~\cite{rei95,gis97}, nanotubes~\cite{eleanor,haifa}, and 
3D quantum dots~\cite{ste00,jon92a} has been thoroughly investigated,
little is known about the non-linear response of 
finite, interacting 2D electronic systems to intense laser fields.  
Is it possible
to observe higher-harmonic (HH) generation, either due to anharmonicities in
the confinement potential, or as a consequence of nonlinearities
in the hydrodynamics of the conduction electrons? Do existing THz sources
(such as free-electron lasers) provide sufficient intensity
to observe HH generation in such systems?  

Alternatively, one may consider weak fields (linear regime),
but high conduction-electron densities. However, most theoretical studies 
of the linear response either consider
the case of low densities of conduction electrons where
the confinement potential can be assumed to be parabolic~\cite{bre90}, or
the other extreme, ``classical" confinement
by infinitely high potential barriers~\cite{scha92}.
How the nature of the optical response changes in realistic systems as 
the density of conduction electrons  is increased 
(so that they explore more and more of the
anharmonic parts of the confinement potential $\vjel$) is not known.
How does the Kohn mode compete 
with other modes of excitation when the HPT is no longer valid? 
 
Last but not least, 
to which extent can a classical model~\cite{fet86} of the (non-)linear
response be adequate in a regime beyond the HPT? Such a 
model would have to account for the hydrodynamics of
the electron fluid.
In the present article 
we address the above questions
within a classical hydrodynamic model, 
and within a quantum-mechanical approach, 
time-dependent density-functional theory~\cite{gro90} (TDDFT).

\begin{figure}
\centerline{\includegraphics{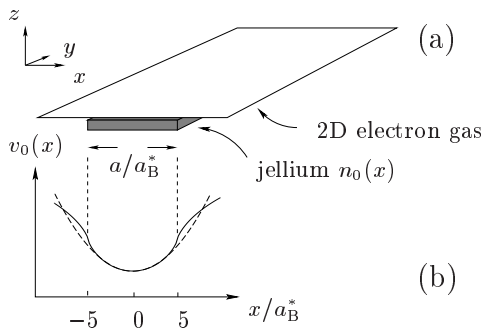}}
\caption{\label{fig:struc} (a) Schematic representation of the system
considered, a 2D electron gas in the $x$-$y$-plane, confined further in the
$x$-direction by a
positively charged (charge density $n_0$), rigid jellium strip oriented along
the $y$-axis, of width $a$. 
A filling
fraction is defined by $\eta = \overline{n}/n_0$.
(b) Electrostatic potential $v_0(x)$
[$\,$arb. un.] of the
jellium charge (solid line)
for $a=10 a_{\rm B}^\ast$ and $r_0 \approx 0.47\au$. Reduced atomic units are
used throughout, $a_{\rm B}^\ast$ is the reduced Bohr radius. 
 Around
 $x\!=\!0$, the confinement potential is harmonic, $\vjel(x)\approx(K/2)\, x^2$
  (dashed line)
 with $K = 8n_0/a$.  For large values of $x$ ($|x| > a/2$), $v_0$ grows
 logarithmically.  }
\end{figure}

{\em Model.} Our model is described in Fig.~\ref{fig:struc}.
We assume that the confinement
potential $\vjel$ is supplied by a 2D rigid positive jellium charge. 
If neutralised
with charge carriers 
this model corresponds  to a 2D metallic strip~\cite{san00}.
It can be regarded as a model of
modulation-doped semiconductor heterostructures 
embedded in a dielectric medium where a layer of dopant charges
corresponds to the positive background. How these, in 
combination with vertical gate voltages, 
can modify the overall confinement potential
is discussed in Ref.~\onlinecite{lie93}. 

In the following we show results for two cases,  wide and
 narrow confinement in the $x$-direction 
($a=100\, a_{\rm B}^\ast$ and $10\, a_{\rm B}^\ast$, respectively); 
corresponding to very shallow confinement in the case of $a=100\, a_{\rm B}^\ast$, 
and very strong confinement  for $a=10 \,a_{\rm B}^\ast$.
In GaAs, the widths correspond to roughly $1\,$$\mu$m and  
$100\,$nm, respectively.
 The
 filling fraction $\eta$ is a parameter ($0 \leq \eta \leq 1$).
 For a given value of $\eta$, the electron charge per unit length
 (in the $y$-direction)
 is taken to be the same in both cases. 
The system is subjected to an electric field $\mathbf{E}(t) = E_x(t)
\hat{\mathbf{e}}_x$ pointing in the $x$-direction, $\hat{\mathbf{e}}_x$. 

{\em Methods.} Due to translational symmetry in the $y$-direction,
the problem reduces    to a one-dimensional self-consistent one, 
of determining the dynamics of the electronic density profile $n(x,t)$ in the potential
\begin{eqnarray}
v([n];x,t)&=&x E_x(t) +v_{\rm xc}([n];x,t) \\
&+&2\int\!\rd x'\,[n(x',t)-n_0(x')]\,\log|x-x'|\,.
\nonumber
\end{eqnarray}
Our quantum-mechanical solution~\footnote{Within the TDDFT, 
the dynamics
is determined by the time-dependent
Kohn-Sham  equations~\cite{gro90}.
They were solved numerically using a
split-operator technique~\cite{cal00},  
a 6th-order finite-difference approximation of the kinetic
energy operator and a 13th-order 
Taylor expansion of the propagator.
}
to this problem relies on
the TDDFT~\cite{gro90};  the exchange-correlation potential $v_{\rm xc}([n];x,t)$ 
was treated in the adiabatic local-density approximation (ALDA)~\cite{yab96}.
The quantum-mechanical wave-packet dynamics was started from the ground state
of the unperturbed system (with $E_x=0$), i.~e., the solution of the
static Kohn-Sham equations~\cite{koh65}.

We have compared our TDDFT results
with a classical approach:
the classical hydrodynamics of the electron fluid (neglecting $v_{\rm xc}$)
was solved in a co-moving Lagrange frame~\cite{mos00}, represented by a layer of infinitely thin rods of width $\rd x$ with initial positions $x$, infinitesimal charges $n(x,0)\rd x$ (per unit-length), 
and velocities $u(x,t)$, evolving according to Newton's law
$\partial_t u(x,t)=-\partial_x v([n];x,t)$
The classical wave-packet dynamics was started from the stationary
solution $n(x)$ of $\partial_x v([n];x,0)=0$.
It was  found by
relaxing an initial Gaussian density profile with additional,
suitably chosen Stokes damping.
Since the static screening length of a 2D electron
gas is considerably larger than the inter-particle distance, 
this classical hydrodynamical approach is expected to work well.
It neglects exchange, correlation, and shell effects.

\begin{figure}
\centerline{\includegraphics{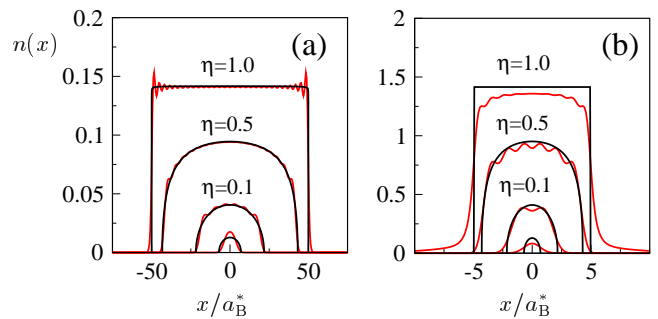}}
\caption[]{\label{fig:clden} 
(a) Classical (black lines) and quantum-mechanical (red lines) ground-state 
charge-density distributions for $a = 100\,a_{\rm B}^\ast$,
for different values of the parameter $\eta$ (charging fraction), 
from $\eta=0.01$ to $\eta = 1$.
(b) Same, but for  $a = 10\,a_{\rm B}^\ast$.  }
\end{figure}

\emph{Ground-state properties.}
Classical and quan\-tum-me\-cha\-ni\-cal
ground-state density profiles
are shown in Fig.~\ref{fig:clden}. 
For the wide system,
the agreement between classical and quantum-mechanical profiles
is satisfactory for most values of $\eta$, except for large $\eta$ (where the quantum-mechanical profile
exhibits Friedel oscillations), 
and for small values of
$\eta$ (where the quantum-mechanical profile
is Gaussian while the classical one is elliptic).
In the narrow system, the discrepancies are larger.
For large $\eta$, 
electron spill-out dominates the quantum-mechanical profile.

\begin{figure}
\centerline{\includegraphics{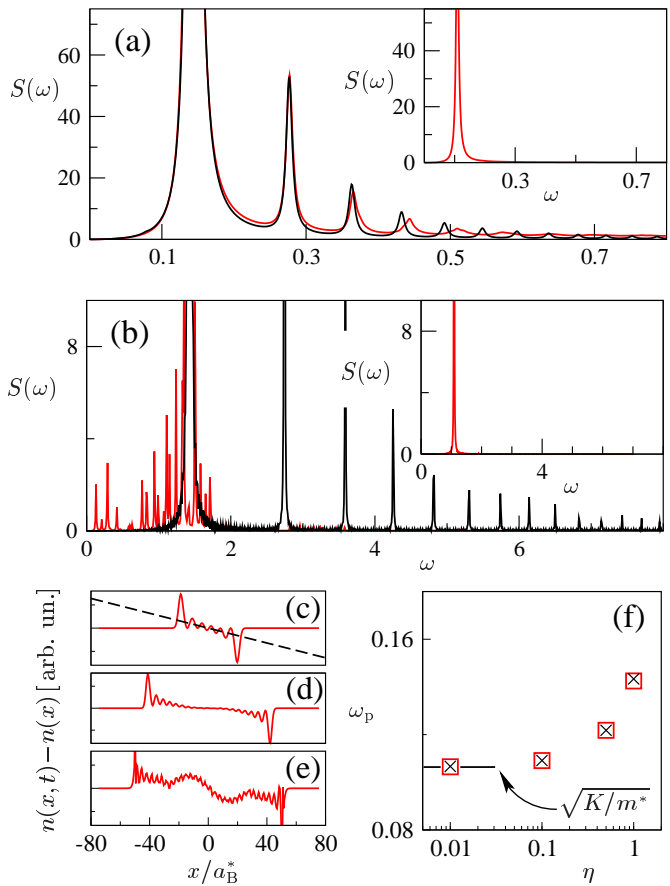}}
\caption[]{\label{fig:linres}
(a) Quantum-mechanical (red lines) and classical
(black lines) linear response for $a=100 \,a_{\rm B}^\ast$ and $\eta = 1$.
Inset: same, but for $\eta=0.1$. 
(b) Same, but for $a=10 \,a_{\rm B}^\ast$ and
$\eta=0.1,1$. (c-e) Snapshots of density profiles for  $a=100 \,a_{\rm B}^\ast$ and
$\eta = 0.1,0.5$ and $1$. For $\eta=0.1$, the profile is dominated
by the Kohn mode 
(dashed line).
(f) Position of the first 
plasmon resonance as a function of $\eta$ for $a = 100\,a_{\rm B}^\ast$.
Shown are classical ($\times$) and quantum-mechanical results ($\Box$).
}
\end{figure}

\emph{Linear response.}
The linear response is obtained by  
applying a low-intensity white-light 
pulse $E_x(t)=E_0 \,\delta(t)$.  A 
value of $E_0=0.001 \au$ was verified to be 
sufficiently small to remain within the regime of linear response,
for the parameters considered here. We have calculated 
the dipolar strength function $S(\omega)=(2\omega/E_0\pi) \mbox{Im}\,
d(\omega)$. Here $d(\omega)$ is the Fourier transform of the dipole moment.

In Figs.~\ref{fig:linres}(a) and (b) our
results for $S(\omega)$ are shown.
For low conduction-electron densities ($\eta=0.1$), 
nearly all dipolar strength is in the Kohn mode
for both the wide and the narrow system, as expected.
The classical and quantum-mechanical strength functions 
are almost indistinguishable [insets of Figs.~\ref{fig:linres}(a) and (b)].
As the filling fraction is increased, higher plasmon modes
develop in the case of the wide system [Fig.~\ref{fig:linres}(a)].
Classical and quantum-mechanical results agree fairly well,
except for large values of $\omega$ where the classical 
plasmon dispersion is found to underestimate the quantum-mechanical result.
The results for the narrow system at $\eta=1$ are very different
[Fig.~\ref{fig:linres}(b)]:
here we observe strong Landau fragmentation~\cite{brack93}
of the main peak; all higher-order 
plasmon modes 
disappear. 
The classical approximation is inadequate in this regime.

\begin{figure}
\centerline{\includegraphics{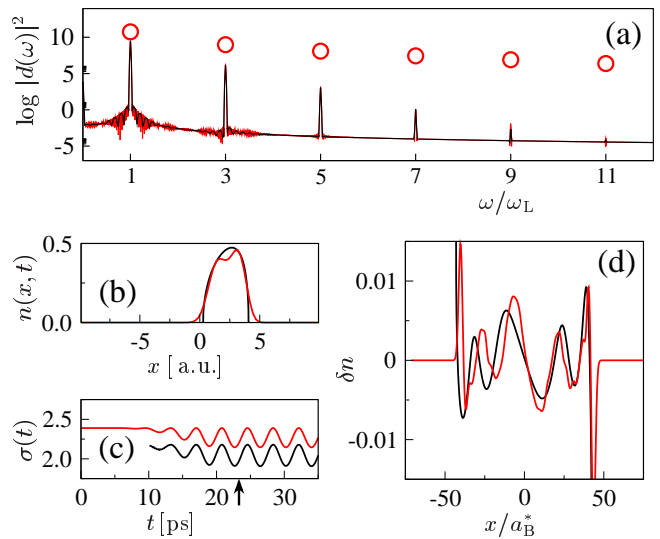}}
\caption[]{\label{fig:hha10}
(a) Non-linear response for $a=10 \,a_{\rm B}^\ast$, $\wl = 0.05\au$
($0.13$ THz),
$\eta = 0.1$, 
and $E_0=3\au$ (intensity $1.38\,10^6\,$W/cm$^2$).
Shown are classical (black lines),
TDDFT results (red lines), and quantum-mechanical independent-particle results (circles).
(b) Density profile for the same parameters
as in (a), but $E_0 = 1\au$ 
 at $t=22.8$ ps.
(c) Time evolution of the width $\sigma$ of this density profile
(the arrow indicates  $t=22.8$ ps).
(d) $\delta n \equiv  n\big(x\!-\!x_{\rm cm}(t),t\big)-n(x)$ for
$a=100 \,a_{\rm B}^\ast$, $\wl = 0.45\au$,
$\eta = 0.5$, and $E_0 = 0.5\au$ (intensity $3.8\,10^4\,$W/cm$^2$),
$t=4.5\,$ps.
}
\end{figure}

What is the spatial profile of the modes observed in Figs.~\ref{fig:linres}(a)
and (b)? The Kohn mode (small $\eta$) is a rigid-shift mode.
As $\eta$ approaches unity in the wide system, 
it evolves into the first plasmon
mode, a combination of a rigid-shift and a hydrodynamic mode
(the higher modes are  expected to be hydrodynamic modes for all values of $\eta$).
These two modes of oscillation correspond to  Goldhaber-Teller
and Steinwedel-Jensen modes in atomic nuclei~\cite{mye77}.
Figs.~\ref{fig:linres}(c-e) show snapshots of quantum-mechanical
density profiles
for three values of $\eta$ in the wide system. 
By virtue of selection rules, $n(x,t)-n(x)$ is  antisymmetric
w.r.t. reflection at $x\!=\!0$.
For small $\eta$,
the rigid-shift mode dominates and $n(x,t)-n(x) \propto -x$
near the origin (note that the rigid-shift profile has the time-dependence
$n(x,t) = n\big(x-x_{\rm cm}(t),0\big)$ where $x_{\rm cm}$ is the
center-of-mass of the profile).
As $\eta$ increases, hydrodynamic modes emerge.
They correspond, approximately,
to standing waves (reflected at $\pm a/2$)
with wave vectors $q=\pi \nu/a$ with $\nu=0,1,\ldots$ and 
 frequencies $\wp^2(q)=2\pi n_0 /(\varepsilon m^*)\,q$
 ($\varepsilon$ is the dielectric constant).
 Here $\wp^2(q)$ is the plasmon dispersion  for 
 a spatially extended 2D electron fluid, the confinement
 is modeled by assuming that the wave-length corresponding
 to $q$ is given by the width $a$.
Fig.~\ref{fig:linres}(f) shows how
the position of the  first ($\nu=1$) plasmon resonance evolves as a function of $\eta$. 
For small $\eta$, the Kohn limit is reached, 
as expected. As $\eta$ is increased,
the position evolves, albeit
not quite to the value $\wp^2\big(\,q\!=\!2\pi/a\,\big) =2\pi^2 n_0 /(a \varepsilon m^*)$.
This is due to the fact that at $\eta=1$, the profile of the density
oscillations is not quite sinusoidal (see also Ref.~\onlinecite{scha92}):
sinusoidal modes do not diagonalise the problem, their
interaction gives rise to a frequency shift.

\emph{Non-linear response.} 
The system was subjected to
an intense monochromatic light wave  of amplitude
$E_0$ and frequency $\omega_{\rm L}$ (switched on slowly,
on the time scale of a few cycles).
The laser intensities were
chosen so as to avoid ionisation of the system,
not exceeding $10^6\,$W/cm$^2$ (well within the
range of standard free-electron lasers).

Fig.~\ref{fig:hha10}(a) shows classical
and quantum-mechanical results 
for the dipolar power spectrum $|d(\omega)|^2$
(see Ref.~\onlinecite{cal00}) in the narrow system.
We observe excellent agreement
between classical and quantum-mechanical results. Further,
we observe HH at odd multiples of $\wl$.
These HH are due to the electrons exploring
the anharmonic potential $v_0$
(c.f. scattering of electrons off the Coulomb potential
in ionised atoms). The parameters ($\eta=0.1$ and $\wl=0.05\au $) were chosen to allow for large excursions of 
$n(x,t)$ into the anharmonic regions of $\vjel$.

Our observations show that confined, interacting  2D electron fluids   do exhibit HH spectra,
albeit not as prominently as in single-electron systems such as atoms
in strong laser fields~\cite{hui92}. 
We surmise that the hydrodynamic modes arising
from the nonlinearities
in the fluid dynamics dampen the center-of-mass motion and its
acceleration,
reducing the intensity of HH.
We have verified that a substantial center-of-mass acceleration
is observed when the non-linear electron-electron interactions
are switched off during the laser pulse, resulting
in HH of considerably larger strength [Fig.~\ref{fig:hha10}(a)].
This implies that 
independent-electron models of 2D interacting electron fluids   are likely
to overestimate the strength of the non-linear response.
Note that our quantum-mechanical results are not sensitive to the
presence/absence of the
exchange-correlation potential, indicating that  $v_{\rm xc}$
has little influence on HH generation.

Following the strong external driving,  the density profile $n(x,t)$ 
moves with the frequency $\wl$, but not rigidly: in the narrow system,
the width of the profile changes periodically (breathing mode), 
as shown in Figs.~\ref{fig:hha10}(b)
and (c). 
This mode dampens the rigid-shift motion of the electron fluid
(c.f. Ref.~\onlinecite{pue01} for a similar effect in
a circular, anharmonic quantum dot).

In the wide system (at $\eta=0.5$),
electron-electron interactions give rise to 
small-amplitude oscillations added to the 
otherwise rigidly moving density profile 
[Fig.~\ref{fig:hha10}(d)].
The time-dependence of the center-of-mass motion is found
to be in good
agreement 
with the classical model.
In the small-amplitude oscillations, by contrast, a
phase shift is observed\footnote{The phase shift is found to increase
 to $\pi$ as $t$ increases.}.
 Finally we emphasise that the selection rules
of the linear case no longer hold
[Fig.~\ref{fig:hha10}(d)].

{\em Conclusions.} 
We have analysed the linear and non-linear
response of confined 2D, interacting electron
fluids to laser light.
Our results may be summarised as follows:
First, non-parabolically confined interacting 2D electron fluids
may exhibit HH spectra under realistic
experimental conditions. HH are due to the
electron fluid exploring anharmonicities in the confinement
potential. It is found that electron-electron interactions dampen
this effect. 
Second, with the exception of small systems at high
electron densities
(where  single-particle excitations interact with the
collective modes giving rise  to considerable Landau fragmentation of
the plasmons), a non-linear classical hydrodynamical
model provides a very good approximation to  the
linear and non-linear response obtained within the TDDFT;
exchange-correlation effects have a negligible influence.
It would be of interest
to ascertain to which extent the non-linear response of
geometrically more complex systems such as nanotubes, quantum rings,
clusters, or C$_{60}$ can be modelled by the classical approach used here.

\end{document}